\documentclass[useAMS,usenatbib]{mn2e}
\usepackage[dvipdfmx]{graphicx}
\usepackage{amsmath,amssymb,natbib}

\begin{document}

\title[Pre-merger localization of eccentric CBC]{Pre-merger localization
of eccentric compact binary coalescences with second-generation
gravitational-wave detector networks}

\author[K. Kyutoku and N. Seto]
{Koutarou Kyutoku$^1$ and Naoki Seto$^2$\\
$^1$Department of Physics, University of Wisconsin-Milwaukee, PO Box
413, Milwaukee, WI 53201, USA\\
$^2$Department of Physics, Kyoto University, Kyoto 606-8502, Japan}

\date{\today}

\maketitle

\begin{abstract}
 We study the possibility of pre-merger localization of eccentric
 compact binary coalescences by second-generation gravitational-wave
 detector networks. Gravitational waves from eccentric binaries can be
 regarded as a sequence of pulses, which are composed of various higher
 harmonic modes than ones with twice the orbital frequency. The higher
 harmonic modes from a very early inspiral phase will not only
 contribute to the signal-to-noise ratio, but also allow us to localize
 the gravitational-wave source before the merger sets in. This is due to
 the fact that high-frequency gravitational waves are essential for the
 source localization via triangulation by ground-based detector
 networks. We found that the single-detector signal-to-noise ratio
 exceeds 5 at 10 min before the merger for a 1.4--$1.4 M_\odot$
 eccentric binary neutron stars at 100 Mpc in optimal cases, and it can
 be localized up to 10 deg$^2$ at half a minute before the merger by a
 four-detector network. We will even be able to achieve $\sim$ 10
 deg$^2$ at 10 min before the merger for a face-on eccentric compact
 binary by a five-detector network.
\end{abstract}


\section{introduction} \label{sec:intro}

Gravitational waves from compact binary coalescences are expected to be
detected in the coming decade by second-generation gravitational-wave
detectors like Advanced LIGO, Advanced Virgo, and KAGRA
\citep{adligo2010,virgo2011,kagra2012}. Gravitational waves will give us
a way to infer masses and spins of binary components, and also equations
of state for supranuclear-density matter if the binary involves a
neutron star. In addition to these intrinsic parameters, extrinsic
parameters such as the luminosity distance and sky position of the
binary can also be studied by gravitational-wave observation. Once we
localize a compact binary coalescence in the sky to the extent that the
host galaxy is determined, the luminosity distance will be known as a
function of the cosmological redshift so that a new distance ladder may
be constructed \citep{schutz1986}. Electromagnetic counterparts of
compact binary coalescences are also important at this stage
\citep{metzger_berger2012,piran_nr2013}, because seeing them is more
advantageous for accurate localization than hearing gravitational waves.

One ambitious goal of gravitational-wave astronomy is to sound a merger
alert before the merger sets in. The benefit of realtime electromagnetic
observation of the compact binary merger will be immeasurable. Nearly
simultaneous detection of gravitational and electromagnetic waves will
give us an opportunity to investigate the propagation speed of
gravitational waves as SN 1987A did for neutrinos
\citep{bionta_etal1987,hirata_etal1987}. Realtime observation of the
merger is also useful to understand electromagnetic radiation
mechanisms. If short-hard gamma-ray bursts are driven by compact binary
coalescences (see \citealt{nakar2007,berger2013} and references therein
for reviews), we will be able to detect prompt emission from the
beginning in various wavelengths for face-on compact binaries. Recent
possible `kilonova' detection supports this binary merger scenario of
the short-hard gamma-ray burst
\citep{berger_fc2013,hotokezaka_ktkssw2013,tanvir_lfhhwt2013}, and we
can reasonably expect simultaneous detection with gravitational
waves. Still, direct connection between the short-hard gamma-ray bursts
and gravitational waves is essential for the robust confirmation of the
scenario. The onset of afterglow will be observed following the prompt
emission. Extended emission observed in the afterglow is one of the most
mysterious feature of the short-hard gamma-ray bursts (e.g.
\citealt{nakamura_knssk2013,veres_meszaros2013}), and detailed
characteristics including its opening angle will be investigated if a
merger alert could be sounded. Other electromagnetic events involving
ultrarelativistic outflows, such as a proposed shock breakout emission
from binary neutron star mergers \citep{kyutoku_is2014}, are also easier
to detect with realtime observation.

Besides the signal-to-noise ratio (SNR), localization of the
gravitational-wave source up to moderate accuracy is necessary for
successful follow-up detection of electromagnetic counterparts
\citep{nissanke_ma2013,ligovirgo2013}, and thus pre-merger localization
should be regarded as a prerequisite for generating merger
alerts. Pointing only known galaxies using available catalogues will
help to detect electromagnetic counterparts particularly in the early
era of gravitational-wave astronomy \citep{nuttall_sutton2010}. Still,
assuming the number density of galaxies to be 0.01 Mpc$^{-3}$, $\sim$ 8
galaxies per deg$^2$ will be found if we are going to observe out to
$\sim$ 200 Mpc. Thus, accurate localization is eagerly desired
irrespective of our follow-up strategy.

Detection and localization of circular binaries on the celestial sphere
prior to the merger are challenging tasks
\citep{ligovirgo2012,cannon_etal2012,ligovirgoswift2012}, whereas such
binaries will be the most frequent sources and are extensively
studied. The reason for this is that a circular binary has the largest
periapsis distance for a fixed value of the semimajor axis, and the
luminosity averaged over an orbit is the lowest among possible
configurations. Furthermore, gravitational-wave frequency is only twice
the orbital frequency within the quadrupole approximation. The
localization of gravitational-wave sources is primarily based on
triangulation via the timing difference between detectors
\citep{fairhurst2011}. Typical separations of ground-based detectors
determine typical timing differences $O(10)$ ms, and thus accurate
localization requires gravitational waves with frequency higher than
$\sim 100$ Hz. But, the gravitational-wave frequency from 1.4--$1.4
M_\odot$ binary neutron stars in a circular orbit does not reach 10 Hz
until $\sim$ 1000 s before the merger, and reaches 100 Hz only at $\sim$
2 s before the merger. As a result, the area of the 90 per cent
localization confidence will be larger than 100 deg$^2$ even 1 s before
the merger for typical events \citep{cannon_etal2012}. Moreover, the SNR
does not substantially accumulate until gravitational-wave frequency
reaches $\sim$ 100 Hz, because each detector is sensitive around 100
Hz. Taking a few to 10 min of latency associated with trigger and alert
generation into account \citep{cannon_etal2012}, it would be challenging
to sound merger alerts with second-generation detector networks even if
we could omit human validation processes. Furthermore, practical
target-of-opportunity observation will involve overheads of telescopes
and satellites as additional latency sources.

The situation could be changed for eccentric compact binary coalescences
(e.g. \citet{oleary_kl2009}), which emit higher mode gravitational waves
even within the quadrupole approximation
\citep{peters_mathews1963}. Event rates of eccentric compact binary
coalescences are conceivably low, and several authors investigated this
possibility. Note that the horizon distance $\sim$ 200 Mpc of a single
detector is not changed drastically by the presence of a finite
eccentricity unless the binary is formed with an extremely short
periapsis distance as we discuss later. \citet{lee_rv2010} studied the
formation of very hard eccentric binaries via tidal-interaction induced
captures in globular clusters focusing on nearly direct collisions. They
found that the local formation rate of binary neutron stars may be a few
tens Gpc$^{-3}$ yr$^{-1}$ depending on globular cluster models. Although
this rate is motivating for ground-based detectors, \citet{east_mlp2013}
pointed out that this might be an order-of-magnitude overestimation
caused by assuming extremely high retention fraction of neutron stars
(see also \citealt{tsang2013}). At the same time, \cite{east_mlp2013}
also pointed out that the cross-section of gravitational-wave induced
captures is larger than that of the tidal-interaction ones by an order
of magnitude. Indeed, whereas the tidal capture requires the initial
periapsis distance to be smaller than $r_\mathrm{p,max} \approx 30$--$40
GM/c^2$ for binary neutron stars with the total mass $M$, where $G$ and
$c$ are the gravitational constant and speed of light, respectively
\citep{lee_rv2010}, the gravitational-wave capture can work up to
$r_\mathrm{p,max} \approx 600$--$700 GM/c^2$ irrespective of the binary
components \citep{oleary_kl2009}. Thus, the merger rate of eccentric
binary neutron stars could be $\sim$ 1 yr$^{-1}$ within the sensitive
volume of second-generation gravitational-wave detectors. Here, we would
not say that such eccentric compact binary coalescences are frequent
enough, because the rate estimation is not settled.\footnote{It is
suggested that dynamical binary-stellar encounters could be another
efficient channel of eccentric compact binary formation
\citep{samsing_mr2014}. It is also suggested that the Kozai mechanism in
hierarchical triples could reduce the time to the merger of inner
compact binaries by several orders of magnitude, and enhance the merger
rate \citep{antonini_perets2012,seto2013,antognini_sta2014}.} Instead,
we consider that they can be possible events and should not be
completely neglected in the gravitational-wave data analysis. Note that
an eccentric binary formed with a very large periapsis distance
$r_\mathrm{p}$ essentially circularizes before gravitational waves are
detected, where the distribution of $r_\mathrm{p}$ is expected to be
flat below $r_\mathrm{p,max}$ from the fact that the capture
cross-section is proportional to $r_\mathrm{p}$.

In this paper, we explore a possibility of realtime detection and
localization of moderately eccentric compact binary coalescences. Here,
we use the word `moderately eccentric' to state that the binary is
eccentric during a detectable inspiral phase, and circularized to a low
value of eccentricity, say $e \lesssim 0.1$, when it reaches the last
stable orbit. In such cases, the remnant of the merger and associated
electromagnetic counterparts are expected to be more similar to those of
circular compact binary coalescences than those of direct collisions. We
only focus on equal-mass, 1.4--$1.4 M_\odot$ binary neutron stars in
this study, and our result depends only weakly on the total mass and
mass ratio once effects of the eccentricity are properly normalized with
respect to the chirp mass. We always assume that binary components are
treated as non-spinning point particles up to the last stable orbit,
while the exact merger is determined by the radius of neutron stars.

The paper is organized as follows. In Section \ref{sec:evolution}, we
briefly summarize the orbital evolution of eccentric binaries. Next, in
Section \ref{sec:analysis}, our formalism to evaluate the SNR, timing
accuracy, and localization accuracy is described. Results are shown in
Section \ref{sec:result}, and Section \ref{sec:discussion} is devoted to
the summary and discussion. We denote masses of each component by $m_1$
and $m_2$. The total mass and reduced mass are written by $M = m_1 +
m_2$ and $\mu = m_1 m_2 / M$, respectively. The distance to the binary
from the earth (or Solar system barycentre) is written by $D$. The
semimajor axis and eccentricity of the binary are denoted by $a$ and
$e$, respectively.

\section{binary evolution} \label{sec:evolution}

We describe the motion of an eccentric binary with $a$ and $e$ in
Newtonian gravity, and consider gravitational radiation in terms of
quadrupole formula. The orbital frequency as the inverse of the period
is given by
\begin{equation}
 f_\mathrm{orb} (a) = \frac{1}{2 \pi} \sqrt{\frac{GM}{a^3}} ,
\end{equation}
and gravitational waves are decomposed into harmonic modes with
frequency
\begin{equation}
 f_n = n f_\mathrm{orb} ,
\end{equation}
where $n \ge 1$ \citep{peters_mathews1963}. The luminosity of each
harmonic mode averaged over an orbit is given by
\begin{equation}
 \frac{\mathrm{d}E_n}{\mathrm{d}t} = - \frac{32}{5} \frac{G^4 M^3
  \mu^2}{c^5 a^5} g (n,e) ,
\end{equation}
where $g(n,e)$ is given by equation 20 of \citet{peters_mathews1963}. We
have
\begin{equation}
\sum_{n=1}^\infty g(n,e) = \frac{1 + (73/24) e^2 + (37/96) e^4}{( 1 -
 e^2 )^{7/2}} = g(e) ,
\end{equation}
and the total luminosity averaged over an orbit
$\mathrm{d}E/\mathrm{d}t$, which governs the orbital evolution, is given
in terms of $g(e)$.

Time evolution of an eccentric binary due to radiation reaction is
handled following \citet{oleary_kl2009} in this study. Specifically, we
use the eccentricity, $e$, as the independent variable, and describe all
quantities including the time to the merger as functions of $e$. The
time evolution of $a$ and $e$ are derived under the assumption of
adiabatic evolution according to \citet{peters1964}, whereas the actual
evolution may be more discontinuous due to the pulse-like nature of
gravitational waves from a highly eccentric binary. This approximation
may require sophistication once we are going to prepare waveform
templates, and we do not expect that it changes the story as far as
pulse-like gravitational-wave trains are emitted.

\begin{figure}
 \includegraphics[width=80mm,clip]{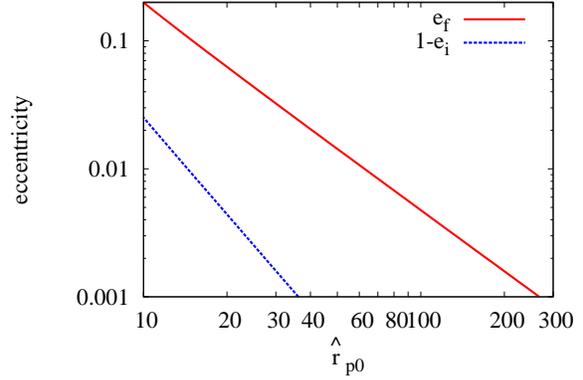} \caption{The eccentricity
 at the last stable orbit $e_\mathrm{p}$ \citep{cutler_kp1994} and
 deviation of initial eccentricity from the unity $1-e_\mathrm{i}$ for a
 gravitational-wave induced capture as functions of the initial
 periapsis distance, $\hat{r}_{\mathrm{p}0}$ \citep{oleary_kl2009}.}
 \label{fig:ecc}
\end{figure}

The periapsis distance is given by $r_\mathrm{p} = a (1-e)$, and we
introduce a normalized periapsis distance
\begin{equation}
 \hat{r}_\mathrm{p} \equiv \frac{a(1-e)}{GM/c^2} ,
\end{equation}
for convenience. We further introduce an initial normalized periapsis
distance at $e=1$ as $\hat{r}_{\mathrm{p}0}$. Although realistic
binaries should be formed with an eccentricity smaller than unity, we
consider the evolution of a binary from $e=1$ as an approximation. This
approximation can be relaxed by evolving a binary from a realistic
initial value at $e_\mathrm{i}<1$, but we do not take this step so that
the formation process is kept unspecified. The change of quantities such
as SNRs associated with setting $e_\mathrm{i} = 1$ is negligible due to
its proximity to unity if we assume gravitational-wave induced captures
(see Fig.~\ref{fig:ecc}).

The periapsis distance evolves in time according to
\citep{oleary_kl2009}
\begin{equation}
 \hat{r}_\mathrm{p} (e) = \hat{r}_{\mathrm{p}0} \kappa (e) ,
\end{equation}
where
\begin{equation}
 \kappa (e) = 2 \left( \frac{304}{425} \right)^{870/2299}
  \frac{e^{12/19}}{1+e} \left( 1 + \frac{121}{304} e^2
			\right)^{870/2299} .
\end{equation}
This implies that
\begin{align}
 a (e) & = \frac{\hat{r}_{\mathrm{p}0} GM}{c^2} \frac{\kappa (e)}{1-e}
 , \\
 f_\mathrm{orb} (e) & = \frac{c^3}{2 \pi G M} \left(
 \frac{1-e}{\hat{r}_{\mathrm{p}0} \kappa (e)} \right)^{3/2} .
\end{align}
The formal time to the merger, at which $e=0$ is achieved, is given by
\begin{align}
 t_\mathrm{m} (e) & = \frac{15}{19} \left( \frac{304}{425}
 \right)^{3480/2299} \frac{\hat{r}_{\mathrm{p}0}^4 G M^2}{c^3 \mu}
 \notag \\
 & \times \int_0^e \frac{e'^{48/19}}{(1-e'^2)^{3/2}} \left( 1 +
 \frac{121}{304} e'^2 \right)^{1181/2299} \frac{\mathrm{d}e'}{e'} .
\end{align}
We use this to denote the time to the merger, whereas we truncate
gravitational radiation at a finite eccentricity $e_\mathrm{f}$ of the
last stable orbit, at which
\begin{equation}
 \hat{r}_{\mathrm{p}} (e_{\mathrm{f}}) = \frac{6+2 e_{\mathrm{f}}}{1+
  e_{\mathrm{f}}}
\end{equation}
is achieved \citep{cutler_kp1994,barack_cutler2004}.

Fig.~\ref{fig:ecc} shows the value of $e_\mathrm{f}$ as a function of
$\hat{r}_{\mathrm{p}0}$ as well as (deviation from the unity of) the
initial eccentricity
\begin{equation}
 e_\mathrm{i} \approx \left( 1 - \frac{85 \pi \mu}{3 \sqrt{2}
		       \hat{r}_{\mathrm{p}0}^{5/2} M} \right)^{1/2}
 \label{eq:iniecc}
\end{equation}
for a gravitational-wave induced capture as an example formation process
\citep{oleary_kl2009}. The eccentricity at the merger, which may
correspond roughly to that at the last stable orbit, is less than 0.1
for $\hat{r}_{\mathrm{p}0} \gtrsim 15$, and thus we expect that the
merger outcome and electromagnetic counterparts are closer to those for
circular cases than those for direct collisions. The time to the merger
from the initial eccentricity $e_\mathrm{i} \approx 1$ is
\begin{align}
 t_\mathrm{m} (e_\mathrm{i}) & \approx \frac{3 c^5}{85 G^3 M^2 \mu} a
 (e_\mathrm{i})^4 \left( 1 - e_\mathrm{i}^2 \right)^{7/2} \\
 & \propto \hat{r}_{\mathrm{p}0}^{21/4} ,
\end{align}
and this is approximately twice the orbital period at the formation,
which is also proportional to $\hat{r}_{\mathrm{p}0}^{21/4}$. Here,
formation via the gravitational-wave induced capture is assumed and
equation \eqref{eq:iniecc} is used. The specific values of $t_\mathrm{m}
(e_\mathrm{i})$ are half an hour, $\sim$ 3 d, and $\sim$ 3 yr for
$\hat{r}_{\mathrm{p}0} = 40$, 100, and 300, respectively. Since the
orbital period, which is proportional to $a^{3/2}$, after $N$ periastron
passages decreases as $1/N^{3/2}$ in a high-eccentricity regime, more
than 10 orbits are always expected as far as the binary is formed with
$e_\mathrm{i} \approx 1$. This may validate the use of the adiabatic
approximation, whereas the discontinuous nature of eccentric binary
evolution should be considered seriously in practice, especially for
binaries formed with small values of $r_{\mathrm{p}0}$.

\section{data analysis} \label{sec:analysis}

We estimate the localization accuracy of a gravitational-wave source
adopting the timing triangulation approximation
\citep{fairhurst2009,grover_ffmrsv2014}. In this approximation, the
source position is reconstructed geometrically using the difference of
gravitational-wave arrival times among detectors in a network. The
localization accuracy is primarily determined by the timing accuracy in
each detector. In this section, we describe computations of `realtime'
SNRs, timing accuracy, and localization accuracy, aiming at pre-merger
localization.

\subsection{signal-to-noise ratio}

One of the most important statistics in gravitational-wave data analysis
is the SNR at a single detector. We always assume that the
matched-filtering analysis is conducted. To circumvent complexities
associated with the sky position and orientation of a binary with
respect to the detector, we express relevant integrals in terms of the
luminosity as an averaged quantity following
\citet{flanagan_hughes1998}, \citet{oleary_kl2009} and
\citet{kocsis_levin2012}. For gravitational waves composed of various
harmonic modes, the averaged (and optimal) SNR is given by
\begin{equation}
 \rho^2 = \sum_{n=1}^\infty \rho_n^2 \; , \; \rho_n^2 = \frac{2G}{5
  \pi^2 c^3 D^2} \int_0^\infty \frac{|\mathrm{d}E_n /
  \mathrm{d}f_n|}{f_n^2 S(f_n)} \mathrm{d}f_n ,
\end{equation}
using the fact that overlaps between different harmonic modes vanish
\citep{barack_cutler2004}. Here, $S(f)$ is one-sided noise power
spectral density of the detector. In this study, we always use the
anticipated noise curve of the zero-detuning, high laser power Advanced
LIGO configuration
(https://dcc.ligo.org/cgi-bin/DocDB/ShowDocument?docid=2974).

To explore the possibility of a realtime merger alert, we extend the SNR
to a function of our time variable, i.e. eccentricity.\footnote{We used
Parseval's theorem to derive the expression in terms of the
luminosity. Although it may seem inappropriate to truncate the
integration with the finite time duration, problems do not occur as long
as the adiabatic approximation is valid.} Using a relation
$(\mathrm{d}E_n / \mathrm{d}f_n) \mathrm{d}f_n = (\mathrm{d}E_n /
\mathrm{d}t) (\mathrm{d}e / \mathrm{d}t)^{-1} \mathrm{d}e$, the realtime
SNR for the \textit{n}th harmonic mode is given by
\begin{equation}
 \rho_n^2 (e) = \frac{G^3 M^2 \mu \hat{r}_{\mathrm{p}0}^2}{c^7 D^2}
  \int_e^1 \frac{g(n,e) u(e)}{n^2 S(f_n (e))} \frac{\mathrm{d}e}{e} ,
\end{equation}
where
\begin{equation}
 u(e) = \frac{192}{95} \left( \frac{304}{425} \right)^{1740/2299}
  (1-e^2)^{1/2} e^{24/19} \left( 1 + \frac{121}{304} e^2
			 \right)^{-559/2299} .
\end{equation}
and $f_n (e) = n f_\mathrm{orb} (e)$. The realtime total SNR is given by
\begin{equation}
 \rho^2 (e) = \sum_{n=1}^\infty \rho_n^2 (e) ,
\end{equation}
and thus, the relation between the SNR $\rho (e)$ and time to the merger
$t_\mathrm{m} (e)$ is given in a parametrized form. In a practical
numerical computation, we truncate the summation at
\begin{equation}
 n_\mathrm{max} (e) = 5 \frac{(1+e)^{1/2}}{(1-e)^{3/2}} ,
\end{equation}
with which 0.1 per cent accuracy is expected to be achieved
\citep{oleary_kl2009}. The realtime SNR obtains its full value at the
last stable orbit as $\rho (e_\mathrm{f}) = \rho$.

Because $n_\mathrm{max} (e)$ diverges at $e=1$, we have to approximate
the integral at $e \approx 1$. Using the fact that gravitational-wave
pulses emitted at $e \approx 1$ are approximately identical, we replace
the gravitational-wave luminosity distribution over harmonic modes at
any value of $e$ larger than a prescribed value $e_\mathrm{cut}$ by that
at $e = e_\mathrm{cut}$, as well as the representative frequency for
each harmonic mode. This gives
\begin{equation}
 \rho^2 (e>e_\mathrm{cut}) \approx \frac{2G}{5 \pi^2 c^3 D^2} \frac{|
  \Delta E |}{g ( e_\mathrm{cut} )} \sum_{n=1}^{n_\mathrm{max}
  (e_\mathrm{cut})} \frac{g ( n , e_\mathrm{cut} )}{f_n^2 (
  e_\mathrm{cut} ) S ( f_n ( e_\mathrm{cut} ) )} ,
\end{equation}
where
\begin{equation}
 \Delta E = - \frac{G M \mu}{2 a ( e_\mathrm{cut} )}
\end{equation}
is the energy radiated between $e=1$ and $e=e_\mathrm{cut}$. We
typically set $e_\mathrm{cut} = 0.995$, and the error associated with
this approximation is at most 0.1 per cent, thus negligible. This is
because the radiated energy is very small in the relevant
regime. Another way to avoid this divergence is to introduce realistic
initial values, $e_\mathrm{i} < 1$.

\subsection{timing accuracy}

For a circular binary, the timing accuracy is often approximated by
inverting a two-dimensional Fisher information matrix for the reference
time $t_0$ and phase $\Phi_0$ (e.g. the coalescence time and phase)
neglecting the correlation with other parameters
\citep{fairhurst2009}. On one hand, it is known that the amplitude and
polarization information of gravitational waves in each detector improve
the localization accuracy \citep{kasliwal_nissanke2013}. On the other
hand, the timing triangulation approximation depends on the Fisher
analysis, which is only accurate at large SNRs. In total, the timing
triangulation approximation with timing accuracy obtained by the
two-dimensional model is found to underestimate the localization
accuracy by a factor of $\sim$ 4 compared to that obtained by fully
Bayesian analysis for all the parameters including the sky position at
realistic SNRs \citep{grover_ffmrsv2014}.

In this study, we evaluate the timing accuracy simply by a
one-dimensional model in which the correlation between the time and all
the other parameters are neglected for both circular and eccentric
binaries. In the terminology of the Fisher matrix, the timing accuracy
is estimated as $\sigma_\mathrm{t} = ( \Gamma_{tt} )^{-1/2}$ using the
Fisher matrix component with respect to the reference time. We find that
the timing accuracy estimated in this one-dimensional model is exactly
the same as the result of a two-dimensional, time-and-phase model for an
eccentric binary formed with $e=1$. This owes to the divergence of the
Fisher matrix component $\Gamma_{\Phi \Phi}$ with respect to the
reference phase in the limit of $e \to 1$. The correlation term
$\Gamma_{t \Phi} ( \Gamma_{tt} \Gamma_{\Phi \Phi} )^{-1/2}$ vanishes in
this limit. Even if the realistic initial eccentricity $e_\mathrm{i}$ is
slightly less than unity, our results are essentially unchanged as the
divergence $\Gamma_{\Phi \Phi} \propto (1-e_\mathrm{i})^{-2}$ is rapid
(see Appendix \ref{app:div}). On another front,
\citet{grover_ffmrsv2014} show for a circular binary that this
approximation gives overestimated accuracy by a factor of $\sim$ 2 than
fully Bayesian analysis, and thus, we assume this as an approximate
upper limit. These facts imply that the contrast between localization
properties of eccentric and circular binaries elucidated by our
one-dimensional model is expected to be a conservative estimation.

The reason of the exquisite accuracy for the reference phase of
eccentric binaries is the different morphology of gravitational
waves. The gravitational waveform of circular binaries is usually called
as a chirp signal, in which only a nearly sinusoidal $n=2$ mode is
emitted with increasing amplitude and frequency throughout their
lifetimes within the quadrupole approximation. The matched filtering for
this case is essentially a problem of matching the phase evolution. By
contrast, gravitational waves from eccentric binaries with $e \approx 1$
are considered as a sequence of nearly identical pulses emitted around
the periapsis, or repeated bursts \citep{kocsis_levin2012}. In this
case, the matched filtering may be considered as a problem of finding
the arrival time of pulse peaks displaced by decreasing orbital
periods. The pulse width is approximately constant at $e \approx 1$,
whereas the orbital period diverges at $e \to 1$. Thus, the orbital
phase is determined accurately to be the periapsis when the pulses are
observed.

The Fisher matrix component for the reference time is obtained using the
fact that the \textit{n}-th harmonic mode gravitational waves are
proportional to $\exp [ \mathrm{i} n ( 2 \pi f_\mathrm{orb} t_0 - \Phi_0
)]$ in the frequency domain. It is given by
\begin{equation}
 \Gamma_{tt} = (2\pi)^2 \sum_{n=1}^\infty \rho_n^2 \overline{f_n^2} ,
\end{equation}
where the \textit{m}th moment of the frequency for \textit{n}th
harmonic, $\overline{f_n^m}$, is defined by\footnote{We separate
$1/\rho_n^2$ in $\overline{f^m_n}$ and $\rho_n^2$ to match the existing
study for a circular binary.} \citep{fairhurst2009,fairhurst2011}
\begin{equation}
 \overline{f_n^m} \equiv \frac{1}{\rho_n^2} \frac{2G}{5 \pi^2 c^3 D^2}
  \int_0^\infty \frac{f_n^m |\mathrm{d}E_n / \mathrm{d}f_n|}{f_n^2
  S(f_n)} \mathrm{d}f_n .
\end{equation}
Similarly to the SNR, the realtime \textit{m}th moment of the frequency
for \textit{n}th harmonic is defined by
\begin{equation}
 \overline{f_n^m} (e) = \frac{1}{\rho_n^2 (e)} \frac{G^3 M^2 \mu
  \hat{r}_{\mathrm{p}0}^2}{c^7 D^2} \int_e^1 \frac{f_n^m (e) g(n,e)
  u(e)}{n^2 S(f_n (e))} \frac{\mathrm{d}e}{e} ,
\end{equation}
and the realtime timing accuracy is expressed as
\begin{equation}
 \sigma_\mathrm{t} (e) = \frac{1}{2\pi} \sqrt{\frac{1}{\sum_n \rho_n^2
  (e) \overline{f_n^2} (e)}} .
\end{equation}
We again truncate the summation at $n_\mathrm{max} (e)$. The relation
between the timing accuracy $\sigma_\mathrm{t} (e)$ and time to the
merger $t_\mathrm{m} (e)$ is given in a parametrized form in a similar
manner to that between the SNR and timing accuracy.

As already noted, the caveat of this model is that it relies on the
Fisher analysis, which breaks down at the low SNR regime. Unfortunately,
we are always interested in the low SNR regime as far as our aim is to
sound a merger alert as early as possible. Comparisons with the fully
Bayesian analysis are conducted in \citet{grover_ffmrsv2014} for a
circular binary, but they have never been done for an eccentric
binary. Furthermore, the validity in the very low SNR regime required
for pre-merger localization is not checked even for circular
binaries. Thus, the Fisher analysis could be a poorer approximation in
realistic data analysis for pre-merger localization of eccentric
binaries than expected from the previous study of a circular binary. We
left more quantitative analysis such as the fully Bayesian one for the
future study.

\subsection{localization}

We follow \citet{fairhurst2011} to estimate localization ability of a
detector network. Localization accuracy depends on a particular
configuration of the detector network. In this study, we focus on a
four-detector network composed of Advanced LIGO Hanford, Advanced LIGO
Livingston, Advanced Virgo, and KAGRA, whereas the expression below is
not limited to this particular network. We also consider a five-detector
network composed of the mentioned four detectors and LIGO India. Their
locations are taken from \citet{schutz2011}. The antenna pattern
functions and noise curves will be different among the detectors, but we
adopt the same value of $\sigma_\mathrm{t} (e)$ at all the detectors for
simplicity. While we have not critically assessed the deviation from
realistic values due to this approximation, we believe that our result
is sufficient to elucidate the effect of eccentricity
semiquantitatively, partly because $\sigma_\mathrm{t} (e)$ is derived by
averaging over the sky position and orientation of the binary.

Once locations of detectors and timing accuracy at them are determined,
the inverse covariance matrix associated with the probability
distribution of the sky position is given by \citep{fairhurst2011}
\begin{equation}
 M_{ij} = \frac{1}{\sum_I (1/\sigma_I^2)} \sum_{I,J} \frac{D_{i,IJ}
  D_{j,IJ}}{2 c^2 \sigma_I^2 \sigma_J^2} .
\end{equation}
Here, lowercase latin indices denote three-dimensional coordinates, and
we do not restrict the estimated source position to be on the celestial
sphere at this stage. Uppercase latin indices refer to the detectors in
a network. The timing accuracy at the detector \textit{I} is written by
$\sigma_I$, and $D_{i,IJ} \equiv d_{i,I} - d_{i,J}$ where $d_{i,I}$ is
the location of the detector \textit{I}. This expression is valid for
the case in which the timing accuracy depends on the detector, whereas
we adopt common values in this study as stated above. This matrix
$M_{ij}$ has three eigenvalues, and they define the inverse squared
localization error of the sky position in the three-dimensional space.

We would like to restrict the sky position of the source to the
celestial sphere \citep{fairhurst2011}. The localization error at the
sky position $(\theta,\phi)$ is given using the projection matrix
\begin{equation}
 P_i{}^j (\theta,\phi) = \delta_i{}^j - n_i (\theta,\phi) n^j
  (\theta,\phi) ,
\end{equation}
where $n_i (\theta,\phi)$ is the unit vector pointing $(\theta,\phi)$
from the centre of the earth. The projected inverse covariance matrix is
given by
\begin{equation}
 \hat{M}_{ij} (\theta,\phi) = P_i{}^k (\theta,\phi) P_j{}^l
  (\theta,\phi) M_{kl} ,
\end{equation}
and the two non-zero eigenvalues of this matrix, $\sigma_1
(\theta,\phi)$ and $\sigma_2(\theta,\phi)$, determine the localization
accuracy around $(\theta,\phi)$. Finally, the localization error $\Delta
\Omega$ with a probability $p$ is approximately given by
\begin{equation}
 \Delta \Omega ( \theta, \phi ; p ) \approx 2 \pi \sigma_1 ( \theta ,
  \phi ) \sigma_2 ( \theta , \phi ) [ - \ln ( 1 - p ) ] .
\end{equation}
In this study, we search the best and worst localization errors over the
sky position $(\theta,\phi)$ for a given detector network, and also take
the average.

\section{result} \label{sec:result}

\begin{figure}
 \includegraphics[width=80mm,clip]{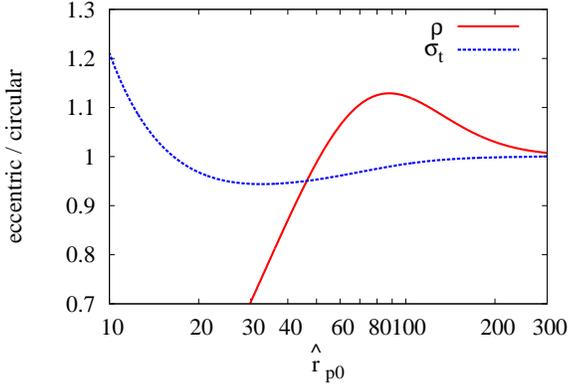} \caption{The dependence of
 $\rho$ and $\sigma_\mathrm{t}$ for a single detector on
 $\hat{r}_{\mathrm{p}0}$. A 1.4--$1.4 M_\odot$ binary is assumed. The
 values are normalized with respect to those of circular binaries, where
 $\sigma_\mathrm{t}$ is defined by $( \Gamma_{tt} )^{-1/2}$.}
 \label{fig:dep}
\end{figure}

\begin{figure}
 \includegraphics[width=80mm,clip]{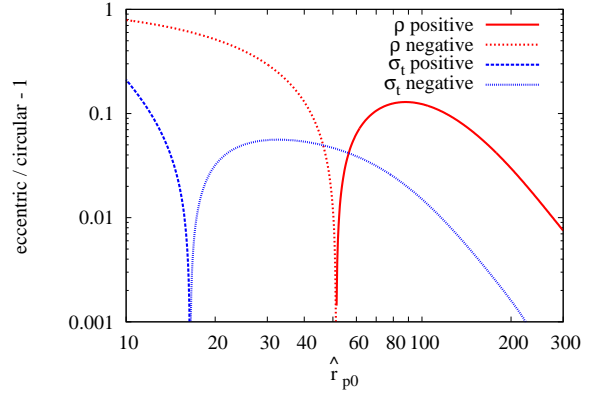} \caption{The relative
 difference of $\rho$ and $\sigma_\mathrm{t}$ for a single detector
 compared to the values of circular binaries. A 1.4--$1.4 M_\odot$
 binary is assumed. The values larger and smaller than those of a
 circular binary are denoted by positive and negative, respectively. For
 better localization, $\rho$ and $\sigma_\mathrm{t}$ should be larger
 ($\rho$ positive) and smaller ($\sigma_\mathrm{t}$ negative),
 respectively.} \label{fig:dif}
\end{figure}

First, we demonstrate the dependence of $\rho$ and $\sigma_\mathrm{t}$
as full values, i.e. those obtained after all the radiation are
detected, on the initial periapsis distance $\hat{r}_{\mathrm{p}0}$ in
Fig.~\ref{fig:dep}. For comparison, all the values are normalized with
respect to those of a circular binary, which take $\rho = 14.2$ and
$\sigma_\mathrm{t} = 0.072$ ms for a binary at $D=100$
Mpc. Fig.~\ref{fig:dif} shows the relative difference in the logarithmic
scale, and representative values for selected eccentric binaries as well
as those for circular binaries are shown in Table \ref{table:data}.

The SNR, $\rho$, takes the largest value at $\hat{r}_{\mathrm{p}0}
\approx 88$. This increase of the SNR at moderately large
$\hat{r}_{\mathrm{p}0}$ is primarily ascribed to the fact that the
radiated energy tends to concentrate around sensitive frequency of a
particular detector configuration. In contrast, the SNR decreases at
smaller values of $\hat{r}_{\mathrm{p}0}$, because the initial orbit is
too close to the last stable orbit and contribution from the inspiral
phase at a large separation is lost. For a large value of
$\hat{r}_{\mathrm{p}0} \gtrsim 300$, the SNR is essentially the same as
that of a circular binary. This figure implies that the horizon distance
is changed only up to a few tens per cent for eccentric binaries except
for those born with very small $\hat{r}_{\mathrm{p}0}$. We do not pay
particular attention to such a weak signal regime in this study.

The timing accuracy for an eccentric binary with $\hat{r}_{\mathrm{p}0}
\gtrsim 15$ is better than that for a circular binary, and it approaches
the value for a circular binary at large $\hat{r}_{\mathrm{p}0}$. We
stress that the values for eccentric and circular binaries agree when we
focus only on the reference time, and the results deviate even in the
limit of $\hat{r}_{\mathrm{p}0} \to \infty$ when we also consider the
reference phase. The timing accuracy becomes worse than that of the
circular binary at $\hat{r}_{\mathrm{p}0} \lesssim 15$ due to the lack
of the SNR, and we do not pay attention to that regime. The best timing
accuracy is achieved at $\hat{r}_{\mathrm{p}0} \approx 32$, whereas the
SNR is smaller by $\sim$ 30 per cent than the circular value.

\begin{table}
 \caption{The SNR $\rho$ and timing accuracy $\sigma_\mathrm{t}$ for a
 single detector of a 1.4--$1.4 M_\odot$ binary at $D=100$ Mpc with
 various values of $\hat{r}_{\mathrm{p}0}$. These for a circular binary
 are also presented as $\hat{r}_{\mathrm{p}0} = \infty$. The timing
 accuracy of a circular binary in the one-dimensional model shown in
 this table is better by 23 per cent than 93.2 $\mu$s obtained in the
 two-dimensional model including the phase.}
 \centering
 \begin{tabular}{c|ccc|ccc|ccc}
  \hline
  $\hat{r}_{\mathrm{p}0}$ & $\rho$ & $\sigma_\mathrm{t}$ ($\mu$s) \\
  \hline
  40 & 12.4 & 67.7 \\
  100 & 16.0 & 70.7 \\
  300 & 14.3 & 71.7 \\
  $\infty$ & 14.2 & 71.8 \\
  \hline
 \end{tabular}
 \label{table:data}
\end{table}

\begin{figure*}
 \begin{tabular}{cc}
  \includegraphics[width=80mm,clip]{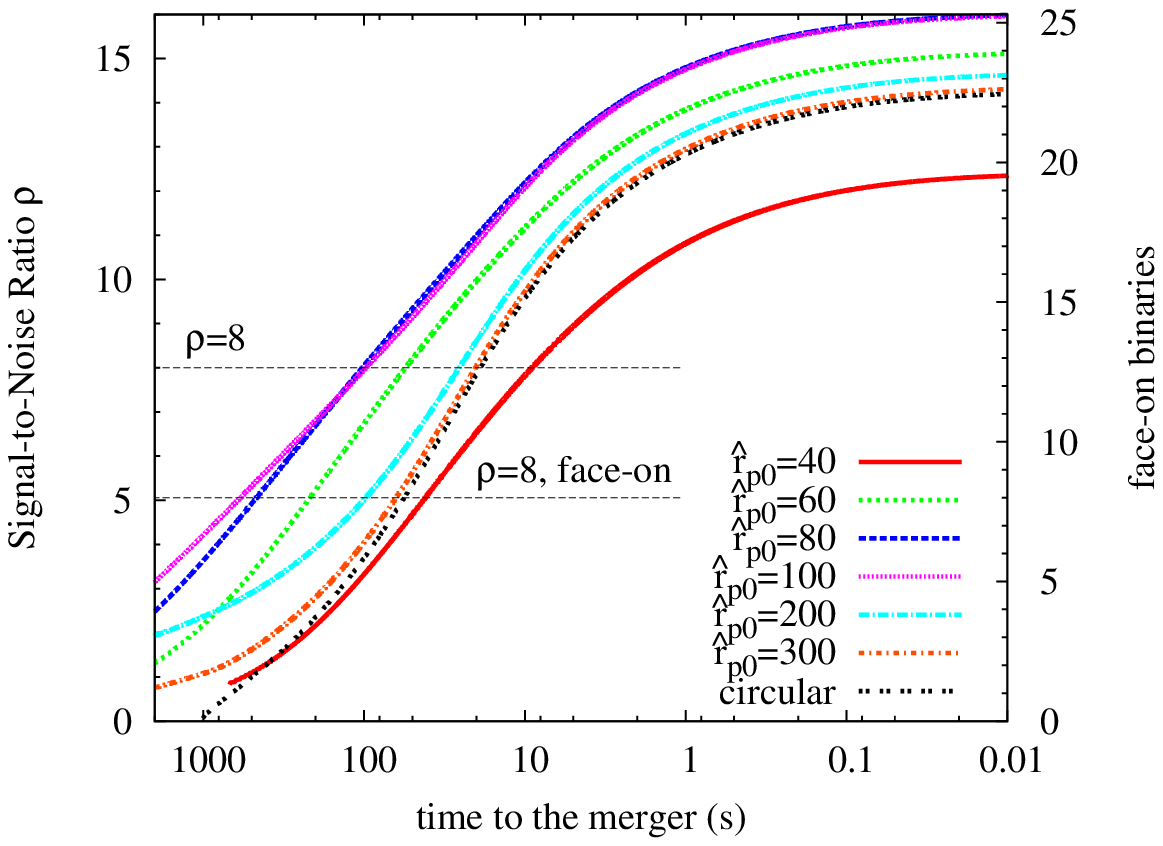} &
  \includegraphics[width=80mm,clip]{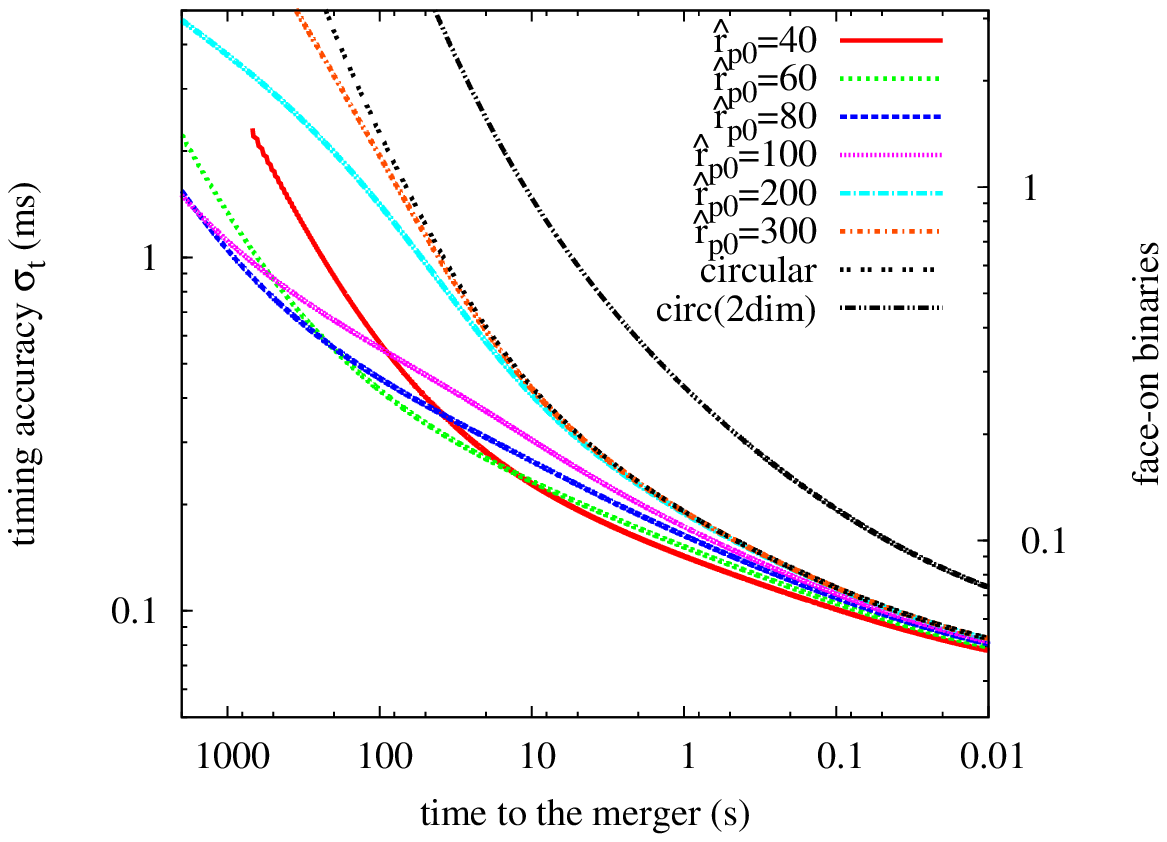}
 \end{tabular}
 \caption{The time evolution of $\rho$ and $\sigma_\mathrm{t}$ for a
 single detector with various values of $\hat{r}_{\mathrm{p}0}$. The
 left and right axes correspond to the orientation-averaged and face-on
 binaries, respectively. All the values are for a 1.4--$1.4 M_\odot$
 binary at $D=100$ Mpc, and averaging over the sky position is always
 performed. Scalings are $\rho \propto D^{-1}$ and $\sigma_\mathrm{t}
 \propto D$. Upper and lower horizontal lines in the left-hand panel
 show $\rho = 8$ for the orientation averaged and face-on cases,
 respectively. We include corresponding curves for the circular
 binaries, and also include timing accuracy obtained by the
 two-dimensional model as `circ(2dim)'. In both plots, curves for
 $\hat{r}_{\mathrm{p}0} = 40$ are artificially truncated at $e =
 e_\mathrm{cut} (=0.995)$, where $t_\mathrm{m} (e_\mathrm{cut}) \approx
 800$ s.} \label{fig:compare}
\end{figure*}

Next, we compare time evolution of $\rho$ and $\sigma_\mathrm{t}$
between eccentric and circular binaries. Fig.~\ref{fig:compare} shows
the time evolution of $\rho$ and $\sigma_\mathrm{t}$ for various values
of $\hat{r}_{\mathrm{p}0}$ as well as those for a circular binary at
$D=100$ Mpc. As an eyeguide, we include horizontal lines which denote
$\rho=8$ both for the orientation-averaged and face-on binaries. Here,
we also show the scale for face-on binaries, because the short-hard
gamma-ray burst should be more relevant for such cases.

The comparison of the SNR evolution shows that eccentric binary
coalescences accumulate SNRs from earlier stages of their lifetimes. A
circular binary at $D=100$ Mpc achieves $\rho = 8$ only at $\sim$ 20 s
before the merger. By contrast, an eccentric binary with
$\hat{r}_{\mathrm{p}0} = 100$ does at $\sim$ 100 s before the merger,
partly due to the large full SNR but mainly due to strong emission at an
early stage of its evolution. If we consider $\rho=5$ or $\rho=8$ of a
face-on binary as a triggering criterion of the merger alert, it is
achieved at $\sim$ 10 min before the merger for
$\hat{r}_{\mathrm{p}0}=100$, whereas it is $\sim$ 1 min for a circular
binary. The rapid SNR accumulation at a few minutes before the merger
occurs approximately for $50 \lesssim \hat{r}_{\mathrm{p}0} \lesssim
200$. For a large value of $\hat{r}_{\mathrm{p}0} \gtrsim 300$, the
accumulation behaviour of the SNR approach asymptotically to that of the
circular case, because the dominant emission channel becomes the $n=2$
mode from a close orbit. For a small value of $\hat{r}_{\mathrm{p}0}
\lesssim 30$, the SNR is not substantial.

The time evolution of timing accuracy also depends on the value of
$\hat{r}_{\mathrm{p}0}$. For a circular binary, $\sigma_\mathrm{t} = 1$
ms is achieved at only $\sim$ 40 s before the merger with $\rho = 6$. By
contrast, an eccentric binary with $\hat{r}_{\mathrm{p}0}=100$ can
achieve the same accuracy at $\sim$ 12 min before the merger. Roughly
speaking, $\sigma_\mathrm{t} = 1$ ms is achieved at a few minutes before
the merger for $50 \lesssim \hat{r}_{\mathrm{p}0} \lesssim 150$. For
virtually all the cases, the timing accuracy as a function of the time
to the merger is always better for an eccentric binary than for a
circular binary, while the SNR can be smaller for a small value of
$\hat{r}_{\mathrm{p}0}$. This reflects the fact that the localization
depends crucially on high-frequency gravitational waves, which a
circular binary does not emit until right before the merger.

\begin{figure}
 \includegraphics[width=80mm,clip]{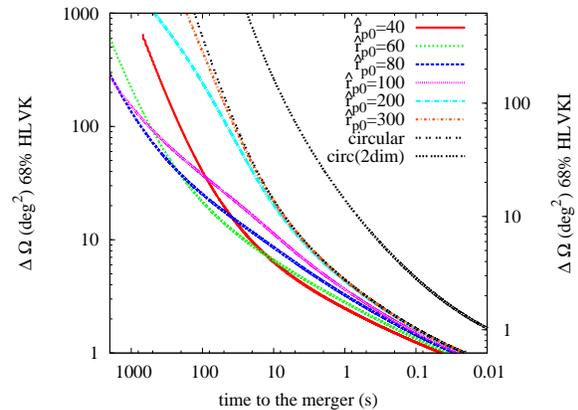} \caption{The time evolution
 of localization accuracy for various values of $\hat{r}_{\mathrm{p}0}$,
 as well as that for a circular binary. The left and right axes are for
 LIGO Hanford-LIGO Livingston-Virgo-KAGRA network (HLVK) and the one
 with LIGO India (HLVKI), respectively. A 1.4--$1.4 M_\odot$ binary is
 assumed to be located at $D=100$ Mpc, and a common value of
 $\sigma_\mathrm{t}$ (and implicitly $\rho$) is adopted for all the
 detectors for simplicity. We include a result obtained by the
 two-dimensional model of timing accuracy for a circular binary, and the
 realistic case may lie within the range between this curve and the one
 for the one-dimensional model \citep{grover_ffmrsv2014}.}
 \label{fig:loc}
\end{figure}

Finally, we compare time evolution of the localization accuracy between
eccentric and circular binaries. Fig.~\ref{fig:loc} shows the evolution
of localization accuracy for various values of $\hat{r}_{\mathrm{p}0}$
as well as that for a circular binary. The localization area is shown as
a sky position averaged value, and the best (worst) case values are
obtained by multiplying 0.4 (3) and 0.5 (2.5) for four- and
five-detector networks considered here, respectively (see Table
\ref{table:net}). Although we specify a detector network, we do not take
directional dependence of the antenna pattern function of each detector
into account, and we take the same values of $\sigma_\mathrm{t}$ at all
the detectors for simplicity. These values depend on the orientation and
sky position of the binary and also orientation of the detector in
reality. Especially, adopting the same values in all the detectors does
not give realistic results for cases in which a binary is located on
blind spots of detectors. Thus, this figure should not be taken too
quantitatively, while we believe that this approximation is sufficient
to demonstrate the localization efficiency.

This figure suggests that eccentric binaries could be localized with 68
per cent confidence up to $\sim$ 10 deg$^2$, typical field of view of
large-area optical telescopes \citep{nissanke_ma2013}, at half a minute
before the merger by a four-detector network. On the other hand, a
circular binary can be localized up to this accuracy only at a few
seconds before the merger. If lower confidence and/or larger
localization error are allowed for the purpose of sounding merger
alerts, more rapid localization is possible.

Taking the expected latency for a gravitational-wave data analysis
pipeline, one interesting value may be those at 10 min before the
merger, at which $\rho=5$, $\sigma_\mathrm{t} = 1$ ms and $\Delta \Omega
\sim$ 100 deg$^2$ at 68 per cent confidence level are achieved. This
should be contrasted with very large localization error for a circular
binary, which will not enable us to declare that the source is localized
in practice. The localization area is approximately halved if we
restrict to 50 per cent confidence area, so that the five-detector
network will be able to localize the source within $\sim$ 10 deg$^2$
even at 10 min before the merger.

Overheads of electromagnetic instruments would determine the required
time to the merger if the latency associated with gravitational-wave
data analysis could be reduced \citep{cannon_etal2012}. They depend on
particular instruments, and one possibility may be 1 min which can be
achieved with robotic telescopes such as Robotic Transient Search
Experiment \citep{rykoff_etal2005}. The SNR exceeds 8 for various values
of $\hat{r}_{\mathrm{p}0}$, and the localization accuracy could be
$\sim$ 10 deg$^2$ by the four-detector network with slightly lower
confidence level than 68 per cent. This level of accuracy is expected
only a few seconds before the merger for a circular binary. Although
this assumption is too optimistic to be realistic, this reminds us that
reducing the latency could have a practical impact on pre-merger
localization accuracy of eccentric binaries.

The localization accuracy is better for a face-on binary, which may be a
progenitor of an observable short-hard gamma-ray burst. The localization
area for an eccentric binary is sufficient for typical gamma-ray
satellites, such as \textit{Fermi} Large Area Telescope (LAT), when
$\rho=8$ is achieved for a face-on one. This occurs earlier by an order
of magnitude for an eccentric binary than for a circular binary (see
Fig.~\ref{fig:compare}). Still, taking not very rapid $\sim$ 0.2 deg
s$^{-1}$ slew speed of \textit{Fermi}-LAT into account, a gamma-ray
satellite capable of rapid timing-of-opportunity observation will be
invaluable even if the sensitivity is inferior to current ones, because
gamma-ray bursts at $O(100)$ Mpc should be extremely bright. The Large
Size Telescopes of the Cherenkov Telescope Array will be able to perform
rapid follow-up to observe more energetic photons at the GeV--TeV range,
which may be able to distinguish mechanisms of the extended emission
\citep{veres_meszaros2013}.

\begin{table}
 \caption{Localization accuracy $\Delta \Omega$ in deg$^2$ for various
 network configurations when $\sigma_\mathrm{t}$ is 1 ms at all the
 detectors. Recall that the localization accuracy scales as
 $\sigma_\mathrm{t}^2 [ - \ln (1-p) ]$ according to the confidence
 $p$. H, L, V, K, and I denote LIGO Hanford, LIGO Livingston, Virgo,
 KAGRA, and LIGO India, respectively. Detector locations are taken from
 \citet{schutz2011}. The `average' and `worst' values for HLV are
 not shown, because the three detectors form a plane and localization
 accuracy is degraded by a factor of $1/\cos \Theta$ where $\Theta$ is
 the angle from the normal to this plane \citep{fairhurst2011}.}
 \centering
 \begin{tabular}{c|cccc}
  \hline
  & HLV & HLVK & HLVI & HLVKI \\
  \hline
  68 per cent best & 160 & 47 & 58 & 34 \\
  68 per cent average & --- & 120 & 120 & 76 \\
  68 per cent worst & --- & 330 & 360 & 190 \\
  \hline
  90 per cent best & 310 & 96 & 120 & 69 \\
  90 per cent average & --- & 250 & 240 & 150 \\
  90 per cent worst & --- & 660 & 720 & 380 \\
  \hline
 \end{tabular}
 \label{table:net}
\end{table}

\section{summary and discussion} \label{sec:discussion}

We showed that gravitational waves from moderately eccentric compact
binary coalescences will not only accumulate a substantial SNR at a few
to 10 min before the merger, but also allow us to determine the sky
position before the merger to the extent that we never expect for
circular compact binary coalescences. This unique property owes to
higher mode gravitational waves emitted during the eccentric inspiral
phase. The pre-merger localization could give us a way to observe the
whole merger process and associated electromagnetic counterparts
provided that the sum of the latency of sounding merger alerts and
overheads of follow-up observation required by telescopes and satellites
can be sufficiently reduced.

The gravitational radiation model adopted in this study is derived by
the quadrupole formula with Newtonian two-point orbital dynamics
\citep{peters_mathews1963,peters1964}. Post-Newtonian corrections such
as the periastron precession and spin-orbit interaction have to be taken
into account for more quantitative study and actual data analysis
\citep{brown_zimmerman2010,huerta_brown2013}, although theoretical
templates of gravitational waves from eccentric compact binary
coalescences are not developed very much compared to those for circular
ones (e.g. \citealt{gopakumar_schafer2011,tessmer_schafer2011}). Tidal
effects due to the finite size of neutron stars, such as the excitation
of \textit{f}-mode oscillation, will also become important at some stage
of the orbital evolution
\citep{stephens_ep2011,gold_btbp2012,east_pretorius2012}. We expect that
these issues modify our conclusion only quantitatively, because the
repeated pulse-like waveform at the early stage of inspiral is
conserved. Another assuring point is that the waveform near the merger
is irrelevant to the purpose of the pre-merger localization and merger
alert, and thus higher order post-Newtonian and tidal effects should be
safely neglected.\footnote{But see, e.g.  \citet{loutrel_yp2014} for
possible ``negative'' post-Newtonian terms in modified gravity with
dipolar radiation.} As stated repeatedly, discontinuous orbital
evolution in the high-eccentricity regime will require more careful
consideration.

The timing triangulation approximation adopted in this study relies on
the Fisher analysis, and it inevitably breaks down when we try to sound
a merger alert using low SNR gravitational waves from a binary long
before the merger. A detailed study such as Monte Carlo simulations is
necessary to assess quantitatively the ability to localize the
gravitational-wave source before the merger with low SNR signals. One
important difference of the eccentric waveform from the circular one is
the introduction of a new degree of freedom, which may be expressed as
the angle of the periapsis direction \citep{barack_cutler2004}. The
correlation between this angle and timing accuracy has to be
investigated quantitatively, whereas we expect that the timing accuracy
primarily depends on the pulse-like waveform structure. In addition, a
common value of the timing accuracy (and SNR) is adopted for all the
detectors to derive the localization accuracy, and this approximation
requires improvement.

A challenging task will be the matched filtering analysis for long-term
gravitational-wave pulse sequences (see also
\citet{antonini_mm2014}).\footnote{See \citet{tai_mp2014} for an
alternative approach to detect signals from eccentric compact binary
coalescences.} In a pipeline of initial LIGO-Virgo data analysis,
detector output data are usually split into 256 s segments
\citep{babak_etal2013}. This will be lengthened in a pipeline of
second-generation detectors, which are sensitive at lower frequency. The
required segment length will be, however, much longer for eccentric
binaries than for circular binaries if we try to process all the
available information, because eccentric binaries emit detectable
gravitational waves during their early lives with very low orbital
frequency (see the end of section \ref{sec:evolution}). One way to
overcome this is to develop a longer pipeline specialized to eccentric
compact binary coalescences. The rotation of the earth will not be
negligible with this strategy. Another way is to choose an appropriate
time interval from the entire lifetime of an eccentric binary for a
given length of the segment. The large number of harmonic modes can be a
difficulty in practical construction of template banks without any
dimensional reduction irrespective of the strategy.

The gravitational-wave pulse from an eccentric binary more resembles a
detector glitch than the chirp signal from circular binaries
does. Sequences of pulses may be naturally distinguished from noises by
multidetector coincidences with information of timing consistency, and
it might require human validation processes at least in the early epoch
of second-generation gravitational-wave detectors. If high false alarm
rates are allowed for the purpose of merger alerts to electromagnetic
follow-up observations, omitting time consuming validation processes
might be an option. Another, possibly ambitious, way to reduce effective
latency might be to first carefully analyse gravitational-wave signals
sufficiently before the merger aiming at rejecting false alarms, and
next analyse signals as close to the merger as possible assuming that
the signal is physical with available information from the first stage
of analysis.

\section*{Acknowledgements}

KK is deeply grateful to John L. Friedman for valuable discussions. KK
is supported by JSPS Postdoctoral Fellowship for Research Abroad, and NS
is supported by JSPS (24540269) and MEXT (24103006).

\appendix

\section{Fisher analysis} \label{app:div}

We compute the Fisher information matrix $\Gamma_{\alpha \beta}$ with
respect to $t_0$ and $\Phi_0$, where Greek indices take $t$ for $t_0$ or
$\Phi$ for $\Phi_0$. Using the fact that the \textit{n}th harmonic mode
is proportional to $\exp [ \mathrm{i} n ( 2 \pi f_\mathrm{orb} t_0 -
\Phi_0 )]$ in the frequency domain, we can replace differentiation with
respect to $t_0$ and $\Phi_0$ by multiplication of $2 \pi \mathrm{i}
f_n$ and $- \mathrm{i} n$, respectively
\citep{cutler_flanagan1994}. Components of the Fisher matrix are given
by
\begin{align}
 \Gamma_{tt} & = (2\pi)^2 \sum_{n=1}^\infty \rho_n^2 \overline{f_n^2}
 , \\
 \Gamma_{t \Phi} = \Gamma_{\Phi t} & = - 2\pi \sum_{n=1}^\infty n
 \rho_n^2 \overline{f_n} , \\
 \Gamma_{\Phi \Phi} & = \sum_{n=1}^\infty n^2 \rho_n^2 .
\end{align}
Computing the inverse of this two-dimensional Fisher information matrix,
we obtain the expression of the timing accuracy,
\begin{equation}
 \sigma_\mathrm{t} = \frac{1}{2\pi} \sqrt{\frac{\sum_n n^2 \rho_n^2}{(
  \sum_m \rho_m^2 \overline{f_m^2} ) ( \sum_n n^2 \rho_n^2 ) - ( \sum_n
  n \rho_n^2 \overline{f_m} )^2}} .
\end{equation}
Note that for a circular binary emitting only via the $n=2$ mode, this
expression gives
\begin{equation}
 \sigma_{\mathrm{t},s} = \frac{1}{2 \pi \rho \sigma_{f,s}} ,
\end{equation}
where $\sigma_{f,s}^2 \equiv \overline{f_2^2} - \overline{f_2}^2$ is the
effective bandwidth for single, $n=2$ mode gravitational waves. This
result agrees with the one derived in \citet{fairhurst2009}.

Now, let us confirm the divergence of phase-related components using a
simplified detector model characterized by
\begin{equation}
 \frac{1}{S (f)} = \frac{\delta(f-f_\mathrm{obs})}{S} ,
\end{equation}
where $S$ is a constant. The SNR is given by
\begin{equation}
 \rho^2 \propto \sum_n \int \frac{|\mathrm{d}E_n / \mathrm{d}f_n|}{f_n^2
  S(f_n)} \mathrm{d}f_n ,
\end{equation}
and thus, we have
\begin{equation}
 \rho^2 \propto \sum_n \frac{1}{f_\mathrm{obs}^2 S} \left|
						     \frac{\mathrm{d}E_n}{\mathrm{d}f_n}
						    \right|_{f_n =
 f_\mathrm{obs}} .
\end{equation}
Hereafter, we replace $\sum_n$ by an integral and express the luminosity
in a more convenient form as
\begin{equation}
 \rho^2 \propto \int \left[ \frac{1}{f_\mathrm{obs}^2 S}
		      \frac{\mathrm{d}E_n}{\mathrm{d}t} \left(
							 \frac{\mathrm{d}e}{\mathrm{d}t}
							\right)^{-1}
		      \left( \frac{\mathrm{d}f_n}{\mathrm{d}e}
		      \right)^{-1} \right]_{f_n = f_\mathrm{obs}}
 \mathrm{d}n .
\end{equation}
In the limit of $e \to 1$, we have
\begin{align}
 a & \propto (1-e)^{-1} , \\
 f_\mathrm{orb} & \propto a^{-3/2} \propto (1-e)^{3/2} , \\
 \frac{\mathrm{d}E_n}{\mathrm{d}t} & \propto \frac{g(n,e)}{a^5} \propto
 g(n,e) (1-e)^5
 , \\
 \frac{\mathrm{d}e}{\mathrm{d}t} & \propto \frac{1}{a^4 (1-e)^{5/2}}
 \propto (1-e)^{3/2}
 , \\
 \frac{\mathrm{d}f_n}{\mathrm{d}e} & = n
 \frac{\mathrm{d}f_\mathrm{orb}}{\mathrm{d}e} \propto n (1-e)^{1/2} .
\end{align}
Therefore, the SNR is approximately written as
\begin{equation}
 \rho^2 \propto \int \left[ \frac{g(n,e)}{n} (1-e)^3 \right] \mathrm{d}n
  .
\end{equation}
Now we can relate $n$ and $1-e$ by $f_\mathrm{obs} = n f_\mathrm{orb}
(e)$ as
\begin{equation}
 n \propto (1-e)^{-3/2} \; , \; (1-e) \propto n^{-2/3} ,
\end{equation}
and we can derive \citep{olver1954}
\begin{equation}
 g ( n , e(n) ) \propto n^{4/3}
\end{equation}
at the limit of $n \to \infty$ ($e \to 1$). Finally, we have
\begin{equation}
 \rho^2 \propto \int n^{-5/3} \mathrm{d}n .
\end{equation}
Thus, the integral (or originally the summation) of $\rho^2$ is not
divergent at $e \to 1$, so is $\Gamma_{tt}$. The integral for $\Gamma_{t
\Phi} \propto \sum n \rho_n^2$ diverges but as moderate as $n
(e)^{1/3}$. The integral for $\Gamma_{\Phi \Phi} \propto \sum n^2
\rho_n^2$ is $n (e)^{4/3}$, and thus violent.

\bibliographystyle{mn2e}

\end{document}